\documentclass[conference]{IEEEtran}
\IEEEoverridecommandlockouts
\usepackage{cite}
\usepackage{amsmath,amssymb,amsfonts}
\usepackage{algorithmic}
\usepackage{graphicx}
\usepackage{textcomp}
\usepackage{xcolor}
\usepackage{hyperref}
\hypersetup{
    colorlinks=true,
    linkcolor=blue,
    filecolor=blue,      
    urlcolor=blue,
    pdftitle={JCET2022},
    pdfpagemode=FullScreen,
}
\def\BibTeX{{\rm B\kern-.05em{\sc i\kern-.025em b}\kern-.08em
    T\kern-.1667em\lower.7ex\hbox{E}\kern-.125emX}}
    
\begin{document}

\title{MT-IceNet - A Spatial and Multi-Temporal Deep Learning Model for Arctic Sea Ice Forecasting\\

\thanks{}
}

\author{\IEEEauthorblockN{Sahara Ali}
\IEEEauthorblockA{\textit{University of Maryland Baltimore County} \\
Baltimore, MD, USA \\
sali9@umbc.edu}
\and
\IEEEauthorblockN{Jianwu Wang}
\IEEEauthorblockA{\textit{University of Maryland Baltimore County} \\
Baltimore, MD, USA \\
jianwu@umbc.edu}
}

\maketitle

\begin{abstract}
  Arctic amplification has altered the climate patterns both regionally and globally, resulting in more frequent and more intense extreme weather events in the past few decades. The essential part of Arctic amplification is the unprecedented sea ice loss as demonstrated by satellite observations. Accurately forecasting Arctic sea ice from sub-seasonal to seasonal scales has been a major research question with fundamental challenges at play. In addition to physics-based Earth system models, researchers have been applying multiple statistical and machine learning models for sea ice forecasting. Looking at the potential of data-driven approaches to study sea ice variations, we propose MT-IceNet – a UNet-based spatial and multi-temporal (MT) deep learning model for forecasting Arctic sea ice concentration (SIC). The model uses an encoder-decoder architecture with skip connections and processes multi-temporal input streams to regenerate spatial maps at future timesteps. Using bi-monthly and monthly satellite retrieved sea ice data from NSIDC as well as atmospheric and oceanic variables from ERA5 reanalysis product during 1979-2021, we show that our proposed model provides promising predictive performance for per-pixel SIC forecasting with up to 60\% decrease in prediction error for a lead time of 6 months as compared to its state-of-the-art counterparts. 
\end{abstract}

\section{Introduction}
The Arctic is a region with unique climate features. For instance, in the Arctic, the Sun never rises over the horizon because of which the seasonal variations in polar day and night are extreme. The enormous areas of Arctic ice and snow are responsible for reflecting sunlight back to space which keeps the planet cool and regulates global and regional weather patterns. However, the Arctic sea ice has seen a continuous decline since 1979 and is left half of which it was in 1970. Therefore understanding Arctic Amplification and forecasting sea ice is a key research topic of climate science. It is important to predict fluctuations in the Arctic sea ice by modeling the weather patterns as it can improve our understanding of potential changes facing the global climate.

To study climate change, environmentalists and domain experts rely greatly on dynamic forecasting systems \cite{johnson2019seas5} that are mainly based on coupled Earth System Models. However, over the last few years, researchers have shifted their focus to data driven Artificial Intelligence (AI) approaches like machine learning and deep learning. Since the climate data presents high spatiotemporal correlations, machine learning models have shown promising results in spatiotemporal data mining leading to short and long-term weather forecasting. Machine Learning (ML) can provide valuable tools to tackle climate change. For example, ML approaches can be used to forecast El-Nino events, hurricanes, and ocean eddies and understand the role of greenhouse gases and aerosols on climate trends and events.

Recent works on climate analytics include Convolutional and Recurrent Neural Network \cite{rs9121305} based models and some hybrid modeling approaches like Convolutional LSTM\cite{xingjian2015convolutional, liu2021daily} and GraphCNN \cite{cachay2020graph}. However, due to the unique nature of the problem of forecasting Arctic sea ice, there are several limitations to the existing solutions and multiple challenges. Enlisted below are some of the prevailing challenges in forecasting Arctic sea ice:
\begin{itemize}
    \item Performance versus lead-time trade-off while predicting per-pixel sea ice variations from sub-seasonal (two weeks to three months) to seasonal (three months to two years) scales. 
    \item Inability to capture the annual minimum and maximum peak values of sea-ice in the non-stationary time-series datasets.
    \item Small data problem owing to the availability of only few decades worth of observational data. 
\end{itemize}

In this paper, we propose a modeling framework, MT-IceNet, to tackle the aforementioned challenges with promising results. Our implementation code can be accessed at the Big Data Analytics Lab GitHub repository\footnote{ \href{https://github.com/big-data-lab-umbc/sea-ice-prediction/tree/main/mt-icenet}{github.com/big-data-lab-umbc/sea-ice-prediction/tree/main/mt-icenet}}.

\subsection{Problem Definition}
In all the research works conducted (details in Section II), there has not been a one size fits all solution proposed to tackle the problem of simultaneously detecting, monitoring and predicting sea ice variations. Therefore, in this paper, we propose MT-IceNet – a fast converging UNet-based \cite{ronneberger2015u} spatial and multi-temporal (MT) regression model for forecasting Arctic sea ice concentration (SIC) at sub-seasonal to seasonal scales. More formally, given:

\textbf{Input}
\begin{itemize}
    \item [--] $X1_i$, monthly observational and reanalysis data where $i=[1,5]$ with a rolling window of 12 monthly records equivalent to one year. 
    \item [--] $X2_i$, bi-monthly observational and reanalysis data where $i=[1,5]$ with a rolling window of 24 bi-monthly records equivalent to one year.
\end{itemize}
MT-IceNet learns from past values of atmospheric and oceans variables (details in Table \ref{vartable}), along with past SIC spatial maps to forecast:

\textbf{Output}
\begin{itemize}
    \item [--]  $Y$, monthly per-pixel Sea Ice Concentration (SIC) values at lead time of $N$ months, where $N = [1,6]$.
\end{itemize}

Here, lead time represents future forecasts of SIC values with a lag of one to six months between the input predictors $X$ and outcome predictand $Y.$

\subsection{Contributions}

In light of the aforementioned background information, the main goal of this research is to develop a spatiotemporal deep learning model that forecasts Arctic sea ice concentration (SIC) at future months, given spatial data at multiple sub-seasonal scales i.e. bi-monthly (15 days) and monthly levels.
Our major contributions are: 

\begin{itemize}
    \item We combine reanalysis and observational meteorological data from multiple sources into two self-curated spatiotemporal datasets of uniform geographic grid and multi-temporal resolutions.
    \item We propose MT-IceNet - a spatial and multi-temporal deep learning model that incorporates a multi-stream learning approach for multi-temporal data and forecast sea-ice on a monthly seasonal scale of up to 6 months.
    \item We perform a thorough comparative analysis between MT-IceNet, baseline models and recently proposed SIC prediction models for forecasting Arctic Sea ice at seasonal scale.
\end{itemize}

The rest of the paper is organized as follows. Some of the important related work is reported in Section II. Section III describes the details of our dataset. Our proposed model is presented in Section IV. Section V provides results and analysis of our experimental study and comparative analysis. Finally, we conclude our paper and share future directions in Section VI.  
\begin{table*}[htbp]
    \caption{Variables included in the Dataset}
    \begin{center}
    \label{vartable}
    \begin{tabular}{|c|c|c|c|c|c|}
        \hline
        Variable & Source & Units & Range & Dimensions & Frequency  \\
        \hline
        Sea Ice Concentration & NSIDC & \% per pixel & 0-100 & 448 x 304 & daily\\
        Longwave Radiation & ERA5 & W/m$^2$ & 0-300 & 66 x 360 & hourly\\
        Rain Rate & ERA5 & mm/day & 0-800 & 66 x 360 & daily\\
        Snow Rate & ERA5 & mm/day & 0-200 & 66 x 360 & daily\\
        Sea Surface Temperature & ERA5 & K & 200-350 & 66 x 360 & daily\\
        \hline
        
        \end{tabular}
    \end{center}
\end{table*}
\section{Related Work}
Majority of the recent works on climate analytics either include Convolutional and Recurrent Neural Network based models or some hybrid modeling approaches like Convolutional LSTM\cite{xingjian2015convolutional} and GraphCNN \cite{cachay2020graph}. \cite{rs9121305} proposed a fully data-driven  Long Short-Term Memory (LSTM) model based approach for Arctic Sea-ice forecasting and compared it with a traditional statistical model; they found that the LSTM showed good performance for 1-month sea ice concentration (SIC) prediction, with less than $9 \times 10^6$ $km^{2}$ of average monthly Root Mean Square Error (RMSE) and around $11 \times 10^6$ $ km^{2}$ of mean absolute error during the melting season. \cite{kim2020prediction} developed a 2D-CNN model that takes as input 8 atmospheric predictors to predict SIC with 1 month's lead time. They compared the performance with Random Forest baseline model, achieving RMSE of $5.76 \times 10^6$ $km^{2}$. \cite{liu2021daily} worked on daily prediction of the Arctic Sea Ice Concentration using reanalysis data based on a Convolutional LSTM Network. They proposed a ConvLSTM model to predict SIC for $T$ timestep given $T-1$ and $T-2$ 25 km resolution observational data from National Snow and Ice Data Center (NSIDC) (2008-2018). They compared their model with a 2DCNN model that takes in a spatial map with pixel grids from $T-1$ timestep. Their model achieved an RMSE of $11.2 \times 10^6$ $km^{2}$ as compared to the 2DCNN with RMSE of $13.7 \times 10^6$ $km^{2}$. 

Ensembling is another hybrid modeling approach where outputs from multiple models are combined to improve performance, whereas it also reduces variance and generalization errors. \cite{kim2019satellite} worked on an MLR $+$ DNN ensemble model using Bayesian Model Averaging to predict sea-ice concentrations for the next 10-20 years. They evaluated their model using correlation co-efficient ($R^2$ score) and achieved normalized RMSE of 0.8. \cite{AliLSTM} proposed an attention-based LSTM ensemble that takes in multi-temporal, daily and monthly, data and predicts sea ice extent (SIE) for $T+1$ timestep, achieving an RMSE of $4.9 \times 10^6$  $km^{2}$. To explore the potential of probabilistic modeling approaches for forecasting sea ice and to aid uncertainty quantification, \cite{ali2022benchmarking} performed a thorough comparative analysis of four probabilistic and two baseline machine learning and deep learning models and published benchmarking results for sea ice forecasting for multiple lead times on these models. They evaluated these models performance using RMSE error and $R^2$ scores and reported Gaussian Process Regression (GPR) to achieve the most competent results. 
Our work takes inspiration from IceNet proposed by \cite{andersson2021seasonal}. IceNet is a U-Net \cite{ronneberger2015u} based probabilistic model for seasonal sea-ice forecasting. Their model takes in images as input and forecasts as output Sea Ice Probabilities (SIP) for three classes (open-water region SIC $< 15\%$, ice-edge region $15\% <$ SIC $< 80\%$, and confident ice region SIC $> 80\%$) for next 6 months. Through probabilistic deep learning, they showed their forecasted values to be competent with the physics-based ECMWF seasonal forecast system SEAS5 \cite{johnson2019seas5}. IceNet is pretrained using Coupled Model Intercomparison Project (CMIP6) 2,220 years (1800-2011) simulation data and is fine-tuned on NSIDC's observational data from 1979 to 2011. They evaluated their model performance on observational data from 2012-2017 using integrated ice edge error (IIEE) and binary accuracy (BACC). Following IceNet, \cite{ren2022data} proposed SICNet, based on a Temporal Spatial Attention Module (TSAM) that captures SIC variations for a lead time of 7 to 28 days. They evaluated their work using Mean Absolute Error (MAE), Mean Absolute Percentage Error (MAPE) and BACC. However, there are two major differences in our proposed MT-IceNet, IceNet and SICNet. One, MT-IceNet produces spatial patterns through per-pixel prediction for SIC values contrary to SIP classification of IceNet. Second, MT-IceNet shows promising results in the prediction of SIC on greater lead times i.e. 1 to 6 months whereas SICNet predicts SIC on a weekly i.e. subseasonal scale.
\begin{figure*}[ht!]
  \centering
  \includegraphics[width=\linewidth]{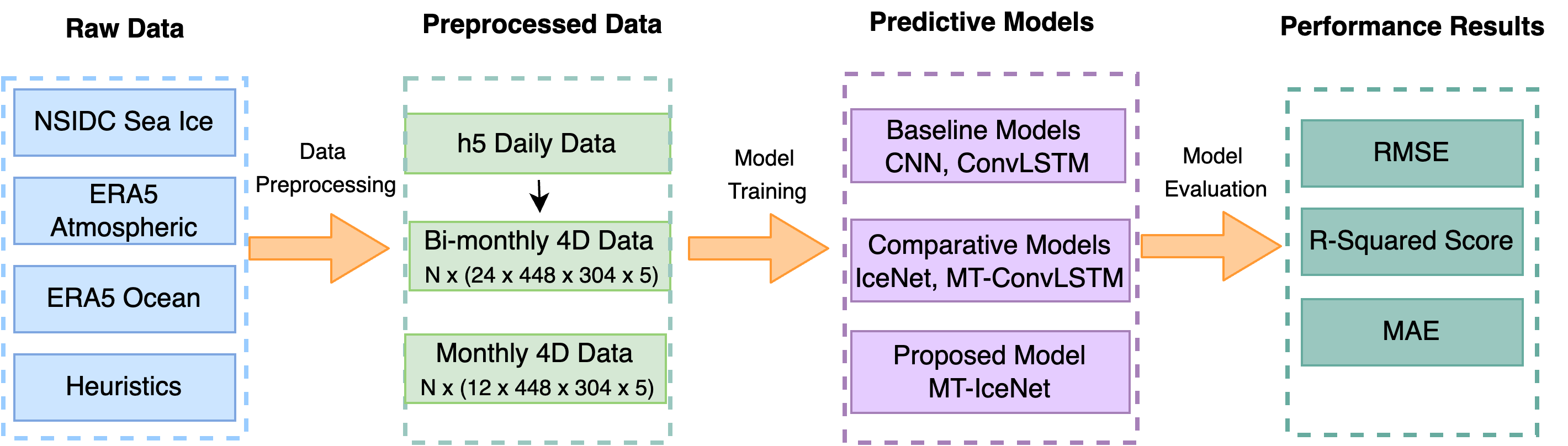}
  \label{fig:pipeline}
  \caption{End-to-end pipeline of our predictive model.}
\end{figure*}

\section{Dataset}

For this study, we use observational sea-ice and reanalysis atmospheric and meteorological data which is available from 1979 till the present. The reanalysis data is available with open access and can be obtained from European Centre for Medium-Range Weather Forecasts (ECMWF) ERA-5 global reanalysis product \cite{ERA5_URL}. Whereas the sea-ice concentration (SIC) values are obtained from Nimbus-7 SSMR and DMSP SSM/I-SSMIS passive microwave data version 1 \cite{Cavalieri1996} provided by the National Snow and Ice Data Center (NSIDC). These variables along with their spatiotemporal resolution details are enlisted in Table \ref{vartable}. 

For the Arctic region, the SIC observational dataset contains an uncertainty of about +-15\% during the summer season due to a high number of melt ponds that can skew the data \cite{Cavalieri1996}. During the winter months, this uncertainty decreases to about +-5\% as the sea ice tends to reach its peak in concentration levels. However, for modeling purposes, this concentration data can be considered as the ground truth.

The inclusion of these variables is based on their causal links with sea ice variations \cite{fdata_arctic_causality} and also based on their physical impact on weather trends in the Arctic. For instance, sea surface temperature provides information on oceanic heat. Similarly, earlier rainfalls during spring trigger earlier Arctic ice and snow melt \cite{Dou,Marcovecchio2021}. Further, as highlighted by \cite{Horvath,huang2019survey}, regional differences in atmospheric pressure cause an increase in Arctic humidity, which in turn enables higher levels of longwave radiation to reach the sea surface. Consequently, this can lead to earlier melting of sea ice. In short, each of the chosen predictors impacts Arctic sea ice through complex oceanic and atmospheric physical interactions.
\subsection{Data Preprocessing}
Each of the chosen data variables come in a different spatial and temporal resolution. For instance, NSIDC provides daily sea-ice concentrations in 25km resolution that is 448$\times$304, whereas the reanalysis data is available in $1^{\circ}$  resolution, i.e. 360$\times$180, as hourly and daily records. For our proposed model, we required 5 dimensional inputs of shape \textit{(samples, timesteps, height, width, features)}. To achieve this, the first step after downloading raw data was to regrid individual $1^{\circ}$ ERA5 variables corresponding to the Arctic geolocation of 90N, 60N, 180E, 180W into the NSIDC polar projections, that is the 25km spatial (448 $\times$ 304) resolution of the sea ice concentration (SIC). To begin with, an empty array with the latitude and longitude geolocation index was created. Next, the values from previous dimensions are interpolated to the new dimensions and stored in the empty grid with new $lat \times lon$ dimensions using the XESMF Python API. The variables acquired in hourly data were aggregated to the daily timescale. After the spatial and temporal rescaling was performed on individual variables, they were combined into a single h5 file with D x H x W x F dimensions. Here $D$ is the total number of days, $H$ is the height of images corresponding to 448 latitude, $W$ represents the width, which corresponds to 304 longitude and $F$ is the number of features which is 5.

\subsubsection{Monthly Data}
To generate the monthly dataset for our model, we averaged the daily 30, 31 or 28 values corresponding to the different months. Special care was taken for leap years, e.g. in case of a leap year, we averaged 29 entries for February. This gave us 504 monthly records. Then we sequentially divided the data into training and testing sets of 408 and 96 months. To reshape the data, a stateless rolling window was applied to the training and testing data, creating 384 samples of 12 months each. Sample one contained months 1-12, sample two contained months 2-13, and the last sample contained months 372-384. Finally we got our training data in the shape $M \times T\times H\times W \times F$, where $M =$ 384 samples, $T =$ 12 months, $H\times W =$  448$\times$304 pixel images and $F =$ 5 features. Similarly, the final shape of the test set was $84\times12 \times448\times304\times5$.
\subsubsection{Bi-Monthly Data}
Deep learning models require a large volume of diverse training data to generalize well on the  unseen test data. However, our monthly dataset is comprised of only 504 records. To counter this small data problem, we generated the second temporal resolution of semi-monthly or bi-monthly data that not only increases the dataset size but also helps our model focus on sub-seasonal patterns. From our previous work \cite{AliLSTM}, we observed high frequency data captures sub-seasonal fluctuations better and in turn helps the model learn the seasonal patterns. For this, similar to aggregating 30 or 31 daily values, we aggregated samples of 15, 16 or 14 daily records depending on the annual months of the year. For example, for January the two bi-monthly records were calculated by taking the average of 15 days and 16 days respectively. Similar to the rolling window applied to monthly data, we applied a 24 timestep rolling window to bi-monthly data to correspond to the same annual cycle as its monthly counterpart. The final dimensions for bi-monthly training and test sets were $384\times24 \times448\times304\times5$ and $84\times24 \times448\times304\times5$ respectively.

\begin{figure*}[ht!]
  \centering
  \includegraphics[width=0.75\textwidth]{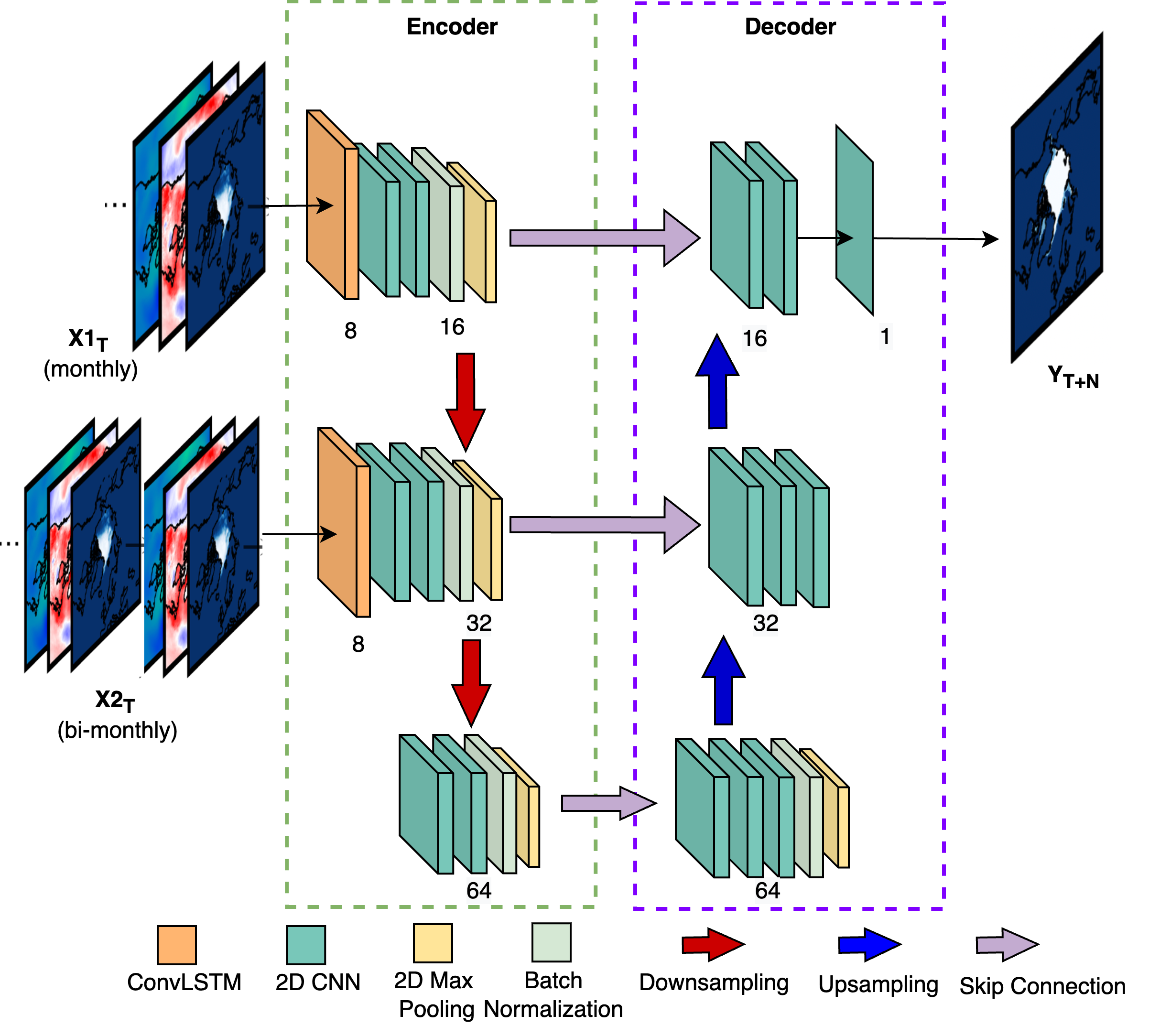}
  \caption{Model Architecture of the Multi-temporal (MT) IceNet Model.}
  \label{fig:archi}
\end{figure*}

\section{Method: MT-IceNet}
Our proposed MT-IceNet model is a UNet-based spatial and multi-temporal (MT) deep learning model for forecasting per-pixel Arctic sea ice concentrations (SIC). The model uses an encoder-decoder architecture with skip connections and multi-temporal data to regenerate spatial maps at future timesteps. As shown in Figure \ref{fig:pipeline}, we started off by downloading the raw data from multiple sources mentioned in Section III. Next, we preprocessed the data to bring it into uniform spatial and temporal resolutions. We then reshaped the data into 5 dimension and sequentially split it into training and testing sets. Sequential splitting is performed to retain the seasonality patterns in the data. We then built our baseline and proposed model MT-IceNet and finally evaluated the performance of all models using the Root Mean Squared Error (RMSE), Mean Absolute Error (MAE) and R-squared ($R^2$) score. The details of our baseline models and constituent blocks of our proposed model are as follows.

\subsection{Baseline Models}
To design our baseline models, we utilized two widely used spatiotemporal deep learning techniques that are, the Convolutional Neural Network (CNN) and Convolutional Long Short Term Memory (ConvLSTM) models. One reason for choosing these two models is that they have been used in previous solutions proposed for this problem. Another reason is that they also work as the constituent blocks of our proposed model MT-IceNet.
\subsubsection{Convolutional Neural Network (CNN)} We designed a simple CNN model with three 2D convolutional layers using the Keras API. Each of these layers was followed by a 2D Max Pooling layer for dimensionality reduction. We flattened the output from the third CNN layer and appended two fully connected (Dense) layers for regression. Finally, we reshaped the output from the final Dense layer into 448 $\times$ 304 to retrieve the predicted spatial maps. The input to this model were mini-batches of 3D tensors of shape $H\times W\times F$ corresponding to $448 \times 304 \times 5$ dimensional images whereas the output was monthly SIC percentage values in the shape of 448 $\times$ 304 images for multiple lead times. To incorporate the lead time in monthly forward predictions, a lag (offset) of 1 to 6 months was created in input features and target SIC values by removing the first 1 to 6 rows from the SIC column and last 1 to 6 rows from the 5 input features. So that a January 1979 data sample would correspond to February 1979 SIC values for a lag of 1 month, a January 1979 data sample would correspond to March 1979 SIC values for a lag of 2 months, and so on. The model was trained six times, each time for a different lag value.
\subsubsection{Convolutional Long Short Term Memory (ConvLSTM)}
We further designed a ConvLSTM model by appending one ConvLSTM layer at the beginning of our baseline CNN model. Since ConvLSTM model requires 4D input tensors, we used the same input train and test datasets generated for our proposed MT-IceNet model for training this ConvLSTM baseline model, that is, $T\times H\times W \times F$. This model was also developed using the Keras API. We trained four ConvLSTM models using the monthly dataset for lead times of 1 to 6 months. To incorporate the lag, an offset was added to train and test sets in a similar manner as done for our CNN baseline model.
\subsection{MT-IceNet Model}
Our proposed Multitemporal (MT)-IceNet is a U-Net based model that comprises two paths of neural network layers, the contractive path that we represent as encoder, and the expansive path represented as the decoder.  The architecture diagram is shown in Figure~\ref{fig:archi}. 
\subsubsection{U-Net}
A U-Net based model was first introduced for image segmentation for bio-medical imagery \cite{ronneberger2015u}. It comprises three constituent blocks; the encoder, the decoder and the bottleneck block that acts as a bridge between both the encoder and decoder. What distinguishes a U-Net architecture from a transformer based model is the use of skip connections between different layers of encoder and decoder. These skip connections provide upsampling layers important features from the downsampling layers that are lost due to the depth of the network.
\subsubsection{Encoder}
Our encoder comprises two downsampling blocks. The first block consists of a ConvLSTM layer that takes in monthly data as input in the shape $T\times H\times W \times F$, here $T = 12$. The output of ConvLSTM layers is passed to two 2D convolution layers, followed by a batch normalization layer and a 2D max pooling layer. The second block follows the same architecture with the difference of input shape. Here, the ConvLSTM layer if given bi-monthly data of the shape  $24\times H\times W \times F$. In every successive layer of the encoder, we increment the output channels by a multiplicative factor of 2, as shown in Figure \ref{fig:archi}. All CNN layers use the same $3\times 3$ kernel size filters whereas the ConvLSTM layer uses a $5\times 5$ kernel. The activation function ReLU is used in all the encoder layers. The encoder part of our model helps learn low-level spatiotemporal dependencies in the data and identifies patterns needed for predicting SIC spatial maps. 
\subsubsection{Decoder}
The purpose of the decoder block is to upsample the low-level features learnt from the data and help reconstruct the spatial map in the same dimension as the input but at a future timestep. Similar to the encoder, the decoder comprises two upsampling blocks. Every block comprises a $2\times2$ upsampling layer using the nearest interpolation method and a $2\times 2$ kernel size filter. The skip connection is built by concatenating the output of each upsampling layer with the output from a corresponding downsampled feature map generated by the encoder, as shown in the Figure~\ref{fig:archi}. Once the outputs are concatenated, they are passed through two 2D convolutional layers. The output channel size of every CNN layer is reduced by a factor of 2 in order to regain the initial input dimension. Finally, a $1\times 1$ convolution with linear activation is applied to the decoder's output to generate the predicted spatial map. 
\subsection{Postprocessing}
We performed two post-processing steps on the predictions generated by our model. Since our predictions correspond to sea ice concentration values that are basically percentage values between 0 to 100, we rescale the values predicted by the model to [0,100] by clipping all predictions less than 0 to 0 and all predictions greater than 100 to 100. This helps interpret the regression results. Further, to help visualize the predictions, we multiplied the predicted spatial maps with a binary land mask. Since we are not interested in land-area predictions, this multiplicative step discards the land area predictions by assigning them zero-weightage while all ocean and water body predictions are retained. This also helps in evaluating the model using the evaluation metrics discussed in Section V. 
\begin{figure*}[ht!]
  \centering
  \includegraphics[width=\textwidth]{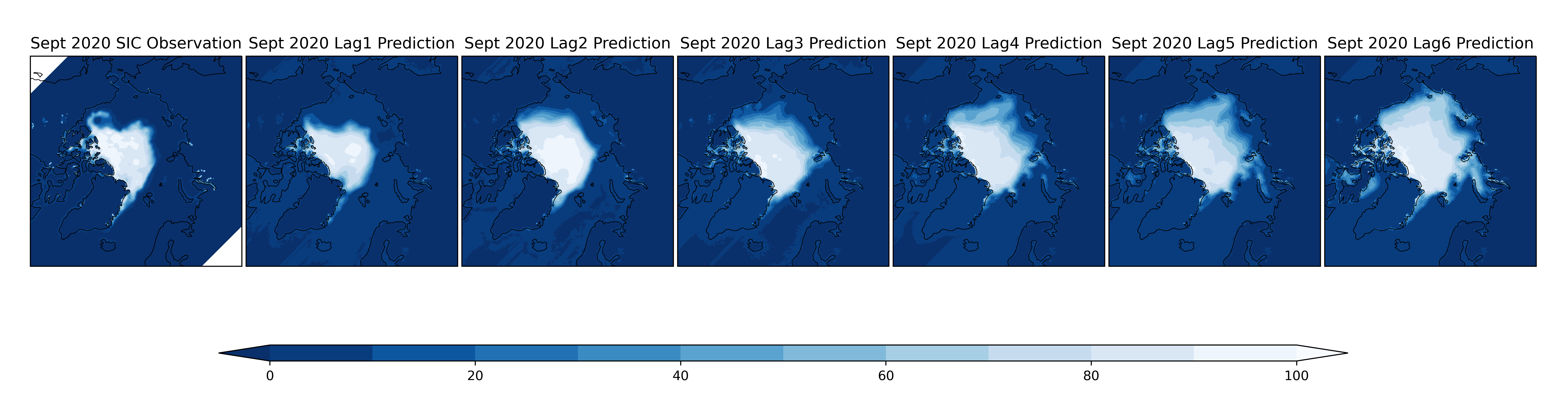}
  \caption{NSIDC Observed Sea Ice Concentration (\%) vs MT-IceNet Predictions for Summer 2020.}
  \label{fig:summer}
\end{figure*}

\begin{figure*}[ht!]
  \centering
  \includegraphics[width=\textwidth]{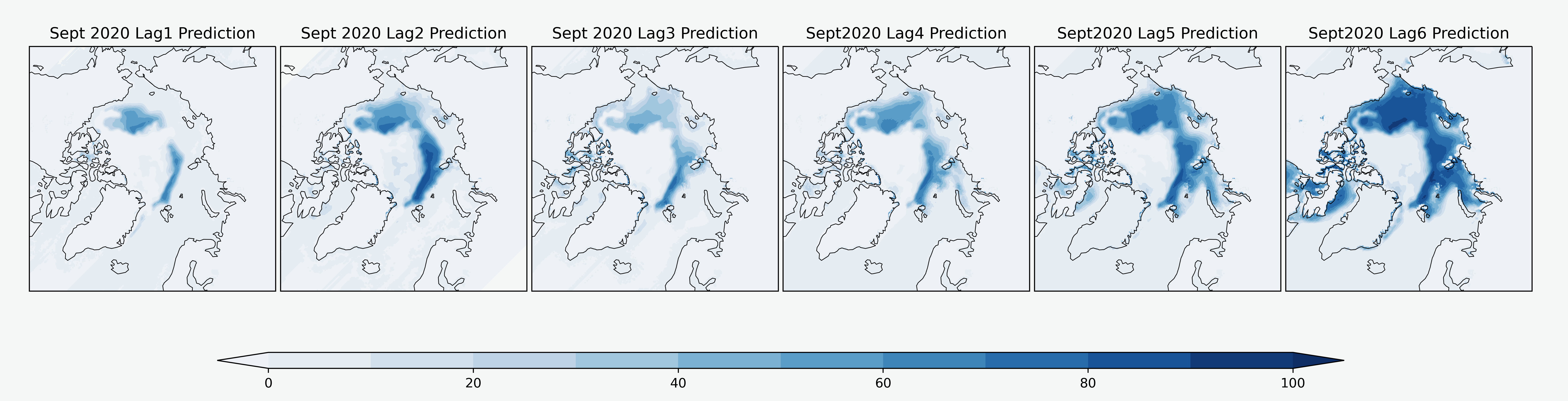}
  \caption{MT-IceNet Summer Prediction difference plots for multiple lead times.}
  \label{fig:sumdiff}
\end{figure*}

\begin{figure*}[ht!]
  \centering
  \includegraphics[width=\textwidth]{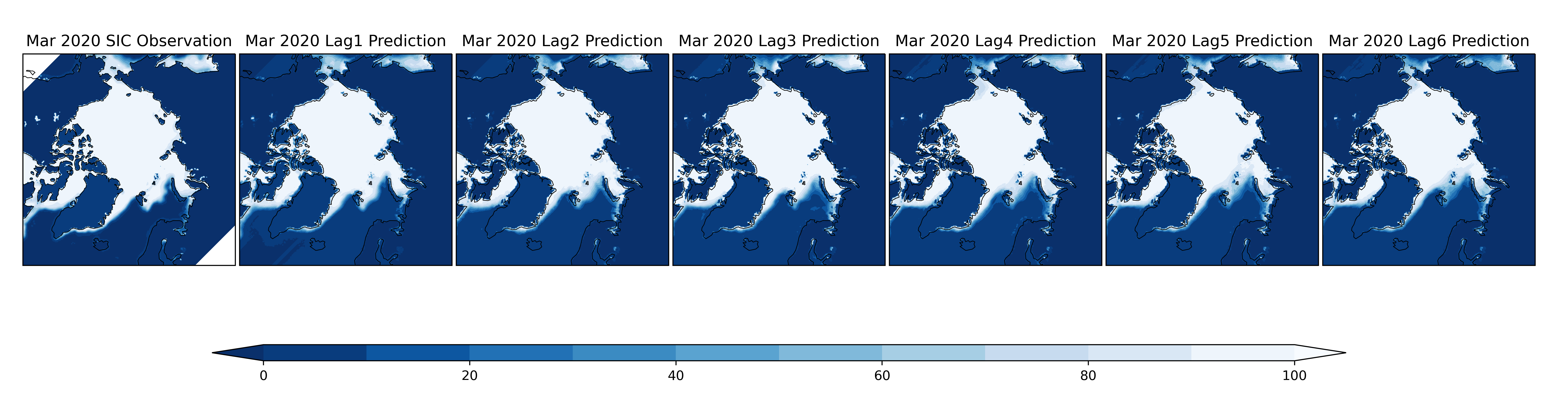}
  \caption{NSIDC Observed Sea Ice Concentration (\%) vs MT-IceNet Predictions for Winter 2020.}
  \label{fig:winter}
\end{figure*}

\begin{figure*}[ht!]
  \centering
  \includegraphics[width=\textwidth]{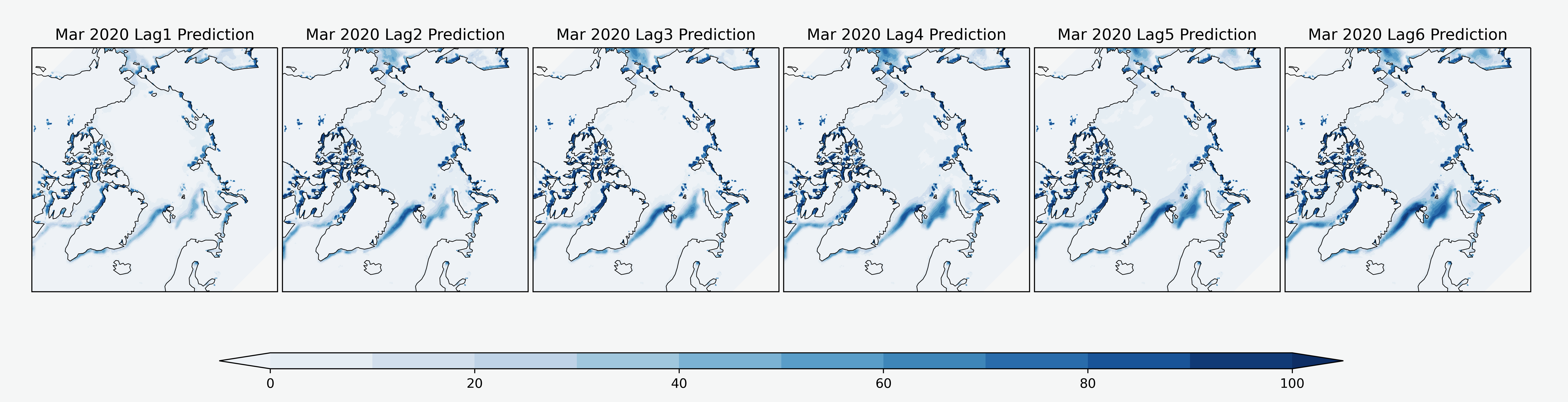}
  \caption{MT-IceNet Winter Prediction difference plots for multiple lead times.}
  \label{fig:windiff}
\end{figure*}
\section{Evaluation}
We first present the experimental setup for our research work in Section V.A. We then move forward to compare our performance with the baseline CNN and ConvLSTM models. We also compared our work with two recently proposed solutions to SIC forecasting using 1) multitask-ConvLSTM method \cite{kim2021multi} and 2) IceNet \cite{andersson2021seasonal}. We present the results of this comparative analysis in Section V.B. Finally we perform the qualitative analysis of our MT-IceNet predictions in Section V.C. 

\begin{figure*}[ht!]
  \centering
  \includegraphics[width=0.85\textwidth]{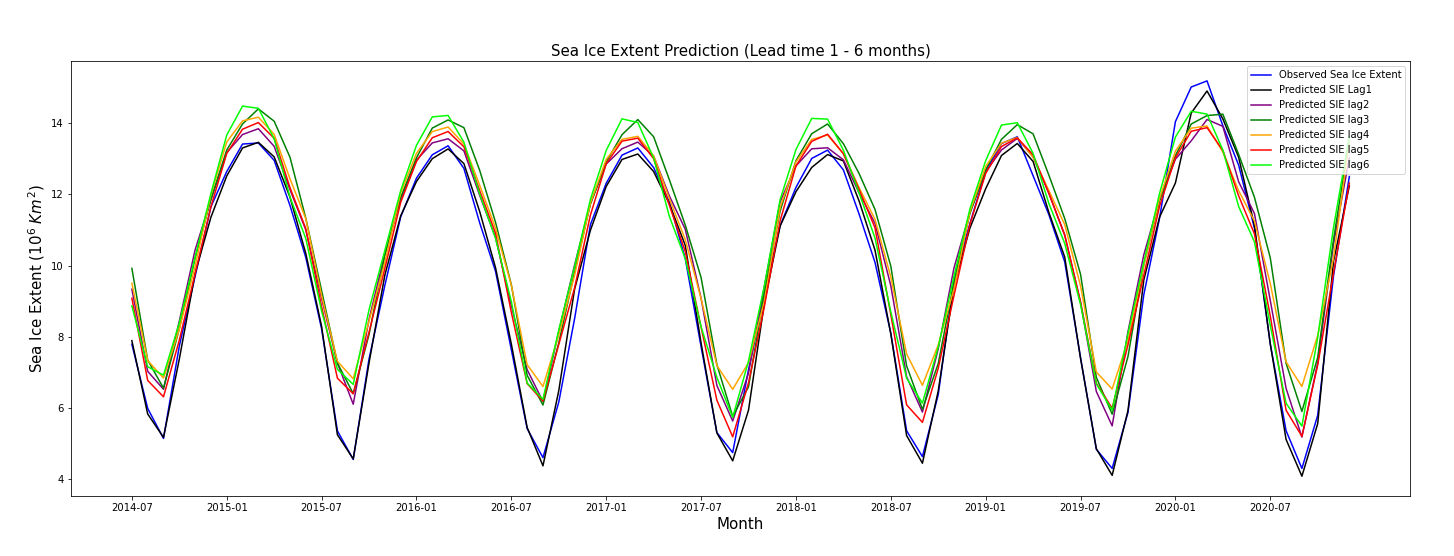}
  \caption{Time-series for derived Sea Ice Extent from MT-IceNet predictions at multiple lead times.}
  \label{fig:ts-sie}
\end{figure*}

\begin{table*}[!ht]
\centering
\caption{RMSE Scores (in \%) for SIC Prediction from multiple models}
\label{tab:rmse-score}
\begin{tabular}{|l|l|l|l|l|l|l|}
\hline
\textbf{Model} & \textbf{One month lag} & \textbf{Two months lag} & \textbf{Three months lag} & \textbf{Four months lag} & \textbf{Five months lag} & \textbf{Six months lag} \\ \hline
CNN                 & 11.34 & 11.75 & 12.51 & 12.71 & 13.03 & 12.72 \\ \hline
ConvLSTM            & 9.12  & 9.45  & 11.00 & 11.07 & 10.26 & 9.89  \\ \hline
Multi-task ConvLSTM\cite{kim2021multi} & 11.73 & 12.02 & 12.61 & 15.79 & 13.12 & 12.86 \\ \hline
IceNet$^\dagger$              & 13.07 & 17.42 & 17.14 & 18.43 & 21.37 & 18.20 \\ \hline
MT-IceNet           & \textbf{5.50}  & \textbf{6.73}  & \textbf{7.61}  & \textbf{7.96}  & \textbf{7.87}  & \textbf{7.77}  \\ \hline
\end{tabular}
\end{table*}

\begin{table*}[!ht]
\centering
\caption{$R^2$ Scores for SIC Prediction from multiple models}
\label{tab:r2-score}
\begin{tabular}{|l|l|l|l|l|l|l|}
\hline
\textbf{Model} & \textbf{One month lag} & \textbf{Two months lag} & \textbf{Three months lag} & \textbf{Four months lag} & \textbf{Five months lag} & \textbf{Six months lag} \\ \hline
CNN                 & 0.838 & 0.826 & 0.802 & 0.796 & 0.786 & 0.796 \\ \hline
ConvLSTM            & 0.892 & 0.885 & 0.839 & 0.840 & 0.860 & 0.870 \\ \hline
Multi-task ConvLSTM\cite{kim2021multi} & 0.823 & 0.813 & 0.793 & 0.673 & 0.773 & 0.781 \\ \hline
IceNet$^\dagger$              & 0.777 & 0.604 & 0.616 & 0.556 & 0.403 & 0.567 \\ \hline
MT-IceNet           & \textbf{0.958} & \textbf{0.929} & \textbf{0.919} & \textbf{0.911} & \textbf{0.913} & \textbf{0.915} \\ \hline
\end{tabular}
\end{table*}

\begin{table*}[!ht]
\centering
\caption{MAE Scores (in \%) for SIC Prediction from multiple models}
\label{tab:mae-score}
\begin{tabular}{|l|l|l|l|l|l|l|}
\hline
\textbf{Model} & \textbf{One month lag} & \textbf{Two months lag} & \textbf{Three months lag} & \textbf{Four months lag} & \textbf{Five months lag} & \textbf{Six months lag} \\ \hline
CNN                 & 3.163 & 3.357 & 3.470 & 3.703 & 3.655  & 4.142 \\ \hline
ConvLSTM            & 2.569 & 2.838 & 3.178 & 3.241 & 3.041  & 3.219 \\ \hline
Multi-task ConvLSTM\cite{kim2021multi} & 3.665 & 3.622 & 3.932 & 5.591 & 4.305  & 4.267 \\ \hline
IceNet$^\dagger$               & 8.564 & 8.749 & 8.448 & 9.536 & 12.277 & 9.157 \\ \hline
MT-IceNet           & \textbf{1.313} & \textbf{1.983} & \textbf{1.962} & \textbf{2.164} & \textbf{2.154}  & \textbf{1.967} \\ \hline
\end{tabular}
\end{table*}

\subsection{Experimental Setup}
All our experiments are performed using the Amazon Web Services (AWS) cloud-based Elastic Compute Cloud (EC2) accelerated computing instances with high frequency 2.5 GHz (base) Intel Xeon Scalable Processor, 32 vCPUs and 64 GBs of GPU memory. The total storage space required for our experiments is around 600 GBs which includes our train and test datasets, model computations and storage and visual illustrations of our results. Our MT-IceNet model is trained using Keras Functional API with a Tensorflow backend and has around 148,000 trainable parameters. This is 99\% less than the 44 million trainable weights of the IceNet\cite{andersson2021seasonal} model. Through a less complex architecture, we also show how simple approaches can generate better results.

We trained our model using Adam optimizer, Root Mean Squared Error (RMSE) loss and trained it on 100 epochs using the Early stopping criteria with a learning rate of 0.0001. Due to the high dimensionality of input data and a limited RAM, we could only process mini-batches of 4 samples each. 

\textbf{Evaluation Metrics:}
We report the RMSE, MAE and $R^2$ performance evaluation scores for our model in Tables~\ref{tab:rmse-score}, \ref{tab:r2-score} \ref{tab:mae-score}. Since it is a spatiotemporal 3D dataset corresponding to latitude and longitude values, we customized the RMSE and MAE metrics for our models evaluation using the following formula:
\begin{equation}
\small RMSE_{SIC} = \sqrt{\frac{\Sigma_{I}\Sigma_{J}{\Big({Y[i,j]-\hat{Y}[i,j]}\Big)^2}}{N}}\label{rmse}
\end{equation}
\begin{equation}
MAE_{SIC} = \frac{\Sigma_{I}\Sigma_{J}\Big|{Y[i,j]-\hat{Y}[i,j]}\Big|}{N}
\label{mae}
\end{equation}
Here, $Y$ represents ground truth while $\hat{Y}$ represents predicted SIC values. $i$ corresponds to 448 latitude, $j$ corresponds 304 longitude values and $N$ represents the total number of test samples. While both, RMSE and MAE, are metrics to calculate error, RMSE gives relatively higher weightage to large error and can help in capturing the variance in error magnitudes.
\begin{equation}
R^{2} = 1 - \frac{RSS}{TSS}
\label{r2}
\end{equation}
We further evaluated our model using the $R^2$ score. As shown in Eq. \ref{r2}, $RSS$ represents the sum of squares of residuals and $TSS$ represents the total sum of squares. Higher $R^2$ score represents better performance. 
\subsection{Comparative Analysis}
For the comparative analysis, we first trained the baseline models CNN and ConvLSTM to predict SIC values for a lead time of 1 to 6 months. We then trained the Multitask-ConvLSTM and IceNet on our dataset to predict SIC values for the same lead times. We are grateful to the authors for providing open-access to their codes on github. Since, IceNet is a computationally expensive classification model that takes in 50 input features and requires 1TB of memory, we customized their models to a light-weight version comprising of only 2 downscaling and 2 upscaling blocks. We further tweaked their output layer by removing the classification layer and replacing it with a regression layer to generate $448 \times 304$ spatial maps at multiple lead times. We refer to this modified version as IceNet$^\dagger$. The multitask-ConvLSTM was proposed to jointly predict per-pixel SIC values and the total sea-ice extent corresponding to the entire spatial region. All these models were trained and evaluated using the same train and test split to have a fair comparison of their performance. In our comparison with other models, we only took into account the SIC prediction and ignored the sea-ice extent values. 
We first analyzed the performance of baseline models for SIC prediction and compared it with our MT-IceNet predictions. 
\subsubsection{Quantitative Analysis}
In Tables \ref{tab:rmse-score} and \ref{tab:mae-score}, it is evident by increasing values of RMSE and MAE scores that both CNN and ConvLSTM have poor predictive performance as compared to MT-IceNet, where our model reportedly decreased the RMSE error by 51\% as compared to CNN, by 40\% as compared to ConvLSTM and by 58\% as compared to IceNet$^\dagger$, for all lead times. The same trend was observed in MAE error where MT-IceNet reduced the MAE error by more than 60\% as compared to its Multi-task ConvLSTM and IceNet counterparts. We further noticed that MT-IceNet has a significantly better $R^2$ score with a notable lead of around 10\% for all lead times as compared to the CNN and ConvLSTM models. The best results have been highlighted in bold in all our result tables.

As seen in Table~\ref{tab:rmse-score}, both multitask-ConvLSTM and IceNet$^\dagger$ take a sharp increase in RMSE score after a lead time time of one month, whereas MT-IceNet still shows a trivial increase in the RMSE scores with only a 2 point increase in RMSE and 1 point increase in MAE from lead time of 1 month to the sixth month. To our surprise, the highest reported errors are from IceNet$^\dagger$ model where the RMSE and MAE errors have significantly higher and $R^2$ scores have very low values for the IceNet$^\dagger$ predictions. It is evident from all three metric results that MT-IceNet outperforms all baseline and recently proposed models for SIC forecasting by showing a promising and persistent predictive performance on greater lead times. The second best performance is achieved by the baseline ConvLSTM model. An interesting observation here is that all models show an improvement in performance after the lead time of 4 months, as evident by the RMSE, $R^2$ and MAE scores. Though this is an interesting finding, the actual cause of this performance improvement is yet to be known. 
\subsubsection{Qualitative Analysis}
To evaluate the quality of our per-pixel predictions, we plot the spatial maps generated by the MT-IceNet model over the Arctic region using Python's cartopy API for geospatial projections. Figures \ref{fig:summer} and \ref{fig:winter} show the forecasted spatial plots where every pixel value lies between [0,100]. Here 100 represents 100\% ice concentration whereas 0 represents the absence of ice in that specific pixel. Since each pixel corresponds to a $25 \times 25$ $km^2$ land area, any value ranging in between 0 to 100 represents the percentage area covered with ice in that region. Looking at Figure \ref{fig:summer}, it is observed that MT-IceNet overpredicts September sea ice at greater lead times which is the trickiest to predict. Nonetheless, our model shows great performance throughout March predictions which is the peak Winter time in the Arctic, as shown in Figure \ref{fig:winter}.

To have a clear identification of regions with incorrect predictions, we plot the differences in the actual SIC observations and the predicted values for multiple lead times, both for Summer and Winter peak months, i.e., September and March, as shown in Figures \ref{fig:sumdiff} and \ref{fig:windiff}. Upon inspecting Figure \ref{fig:sumdiff}, we see that model performs poorly only near the coastal areas of Greenland. For March, we notice that the model underpredicts the sea ice over the coastal areas. This can be considered a minor performance flaw as edge predictions are usually the trickiest for spatiotemporal models. For September predictions, as shown in Figure \ref{fig:sumdiff}, we notice how model overpredicts Summer sea ice at greater lead times. This is due to the concept of seasonal barrier, according to which the seasonality patterns are hard to identify from a distance of more than 3 months. 
Using the SIC values, we calculated the overall sea ice extent for the entire region by calculating the area-weighted sum of the Arctic region using the per-pixel area map provided by NSIDC. We plotted these sea ice extent values as a time-series plot for multiple lead times, as shown in Figure \ref{fig:ts-sie}. We noticed how our model overpredicts summer sea ice and underpredicts winter sea ice at greater lead times. We also noticed the performance improvement in lead times 5 (red) and 6 (lime) where the model predictions once again come closer to the actual observations (blue).
Overall, we did not find any sharp increase or decrease in the SIC model predictions as the lead time increases. This means our model can overcome the performance versus lead-time tradeoff that is faced by most of the models proposed for seasonal predictions.

\section{Conclusions \& Future Work}
In this paper, we presented our work on a spatiotemporal deep learning model that jointly learns from multi-temporal inputs to forecast Arctic sea ice at lead times of 1 to 6 months. Through experiment and ablation study, we showed how our model outperforms the baseline and recent state-of-the-art approaches using a U-Net based architecture by overcoming the small data problem and seasonality barrier challenge. Our MT-IceNet not only outperforms the baseline and other recent work but also shows a consistency in forecasting SIC values at greater lead times. We believe our proposed model can substantially improve our ability in predicting the future Arctic sea ice changes at sub-seasonal to seasonal scales, which is fundamental for forecasting transportation routes, length of open water, resource development, coastal erosion, and threats to Arctic coastal communities and wildlife.

In the future, we plan to extend our work to multi-scale spatiotemporal modeling in order to jointly process fine and coarse resolutions of geolocation information that can be vital in solving similar Earth Science problems. We further plan to incorporate the attention mechanism in our model to identify important contributing factors to the prediction. Lastly, we plan to work on data-driven causal discovery to study variations in Arctic sea ice using spatiotemporal deep learning models.

\section*{Acknowledgement}
This work is supported by NSF grants: CAREER: Big Data Climate Causality (OAC-1942714) and HDR Institute: HARP - Harnessing Data and Model Revolution in the Polar Regions (OAC-2118285). We thank Dr. Yiyi Huang (NASA Langley Research Lab) for her assistance in introducing the dataset.

\bibliographystyle{abbrv}
\bibliography{sample-base}

\end{document}